\documentclass[10pt,twocolumn,a4paper,english]{article}

\usepackage{cite}
\usepackage{graphicx,color}
\usepackage{abstract}
\usepackage{setspace}

\begin{document}

\title{\bf Hadron production measurements to constrain \\ accelerator neutrino beams}




\author{Alexander Korzenev\footnote{Presented at the Neutrino\,2014 conference in Boston, June 2014.
The conference site is http://neutrino2014.bu.edu/} \\
DPNC, University of Geneva, CH-1211 Geneva, Switzerland \\
alexander.korzenev@cern.ch
}

\date{}


\twocolumn[
  \begin{@twocolumnfalse}
    \maketitle
 \renewcommand{\abstractname}{}
\vspace{-1cm}
\begin{abstract}
\noindent {\bf Abstract.}~
A precise prediction of expected neutrino fluxes is required for 
a long-baseline accelerator neutrino experiment. 
The flux is used to measure neutrino cross sections at the near detector, 
while at the far detector it provides an estimate of the expected signal 
for the study of neutrino oscillations. 
In the talk several approaches to constrain the $\nu$ flux are presented.
The first is the traditional one when an interaction chain for 
the neutrino parent hadrons is stored to be weighted later 
with real measurements. 
In this approach differential hadron cross sections are used
which, in turn, are measured in ancillary hadron production experiments.
The approach is certainly model dependent because it requires
an extrapolation to different incident nucleon momenta assuming 
$x_F$ scaling as well as extrapolation between materials having
different atomic numbers.
In the second approach one uses a hadron production yields off 
a real target exploited in the neutrino beamline.
Yields of neutrino parent hadrons are parametrized at the surface of the target, 
thus one avoids to trace the particle interaction history inside the target.
As in the case of the first approach, a dedicated ancillary experiment
is mandatory. 
Recent results from the hadron production experiments
-- NA61/SHINE at CERN (measurements for T2K) and MIPP at Fermilab 
(measurements for NuMI) --  are reviewed.
\end{abstract}
\vspace{0.7cm}
  \end{@twocolumnfalse}
]
\saythanks

\section{Introduction}

A conventional accelerator neutrino beam is formed mainly by the decays
of pions and kaons which, in turn, are generated in interactions of a proton beam  
with a long nuclear target \cite{review,Bonesini}.
Produced in the decays of different mesons at different distances such a beam
is neither monochromatic nor pure in flavor.
In addition to the main component $\nu_\mu$ (or $\bar{\nu}_\mu$) there is 
a significant admixture of different neutrino flavors 
$\bar{\nu}_\mu$,$\nu_e$, $\bar{\nu}_e$ 
(or $\nu_\mu$, $\nu_e$, $\bar{\nu}_e$). 
This admixture is an important source of systematic uncertainty in 
the oscillation analysis.

There are different methods to constrain the $\nu$-beam spectrum.
The most general one assumes a full chain simulation of the neutrino beamline: 
an interaction of primary protons with the target, propagation of secondaries 
through the setup, reinteractions and finally production of neutrinos in the decay.
For each MC event the interaction chain of hadrons is stored to be re-weighted 
later with experimentally measured cross sections \cite{T2K_flux_paper} 
which, in turn, are obtained 
in ancillary hadron production experiments \cite{Bonesini,Boris}.
As a variation of this method the re-weighting of hadron multiplicities could 
be done at a surface of the target, which would take away a dependence of 
the results on an interaction model used for simulation inside the target \cite{NA61_LT}.
An additional constraint for the $\nu$-beam can come from measurements 
by a muon monitor detector placed downstream of the decay tunnel after a beam dump \cite{T2K_NIM,NuMI_MuMon,Poster_0}.
However in general this constraint is rather 'weak' because
momentum of muons is not measured, only intensity and direction.
Besides of these indirect constraints of the neutrino beam 
(constraints on parent or associated particles),
it would be important to mention direct techniques, 
so-called 'in situ' techniques,
which could 
be used to obtain the $\nu$ flux 
by the neutrino experiments themselves.
These techniques are based
on normalization to previously measured cross sections or
normalization to a well theoretically calculated process.
For example, one can use a neutrino-electron process to 
constrain the sum of neutrino and antineutrino fluxes \cite{Poster_1}.
Or so called ``low-$\nu$'' technique which is based on a fact that 
the charged current differential cross section in
the limit of vanishing energy transfer 
is independent of the neutrino energy \cite{Poster_2,Bodek}.
For a high energy neutrino experiment, as NOMAD for instance, one can also 
use an inverse muon decay process to constrain the flux \cite{NOMAD}.
Albeit these direct measurements can assure a precise constraint on 
a certain component of the neutrino beam, 
still input from external dedicated measurements at hadron production experiments
is required 
to provide a full flavor decomposition of the neutrino beam
\cite{LBNE}.

In this talk we consider the T2K/J-PARC and NuMI/FNAL neutrino beamlines,
focusing on recent results of their ancillary hadron production experiments
-- NA61/SHINE at CERN and MIPP at FNAL -- whose measurements are used 
to constrain the neutrino beam.

\begin{figure*}[t!]
\centering
\includegraphics[height=0.33\textwidth]{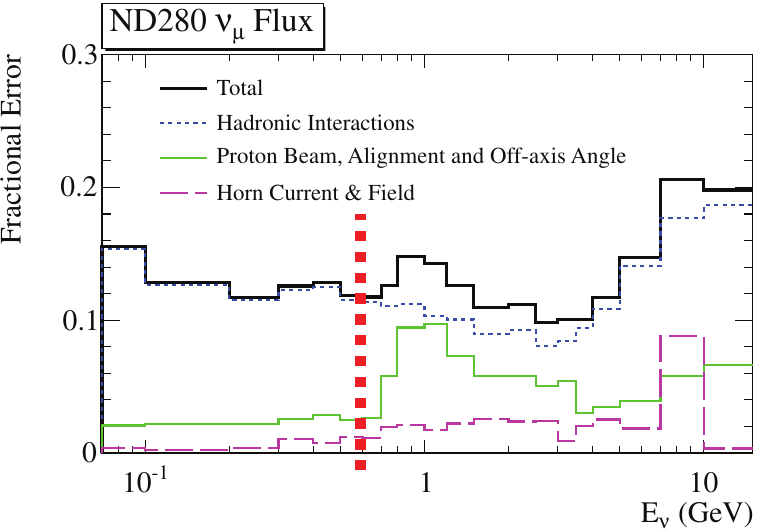}
\hspace{1.5cm}
\includegraphics[height=0.32\textwidth]{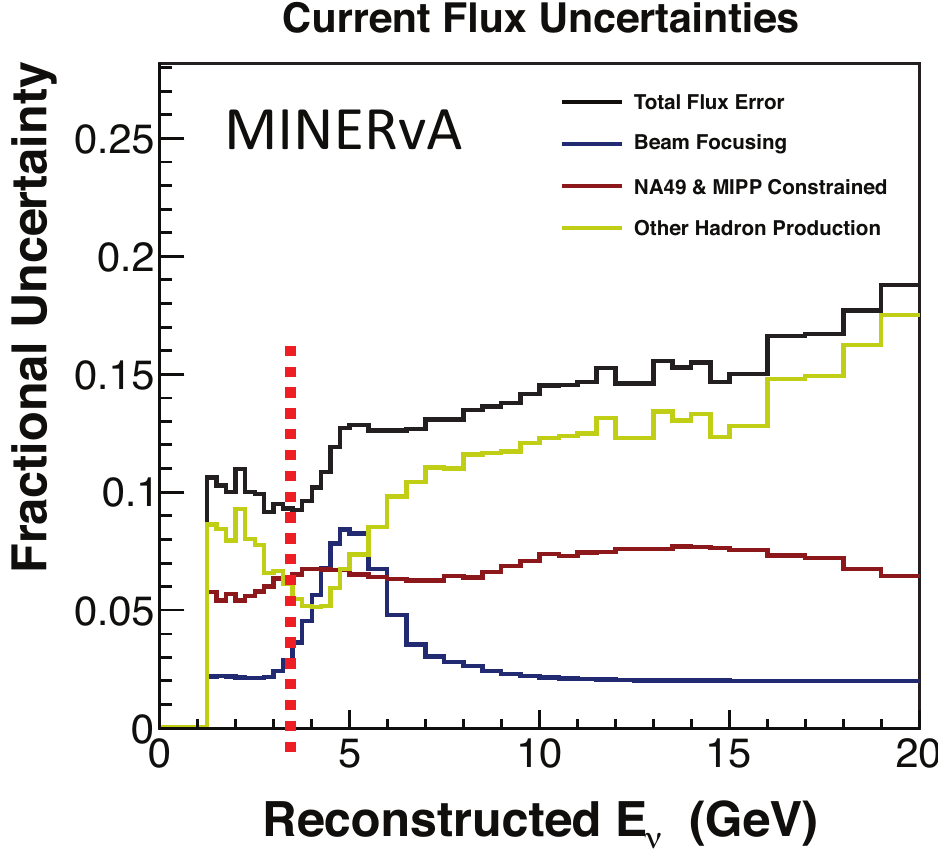}
\caption{Fractional uncertainty for the $\nu_{\mu}$ beam
  calculated for the near detector of T2K \cite{T2K_flux_paper} (left) 
  and for MINERvA \cite{MINERvA_uncert_Harris} (right) 
  as a function of the $\nu_{\mu}$ energy, $E_\nu$,
  are shown. Vertical lines at both figures show a peak value of 
  the $E_\nu$ spectrum.
}
\label{fig:Uncert}
\end{figure*}

\section{The T2K beamline}

The neutrino beam for T2K is generated by the \mbox{J-PARC}
high intensity 31~GeV proton beam interacting with
a 90~cm long graphite target \cite{T2K_NIM}. 
Produced particles are focused by three magnetic horns  
which are followed by the 100 m long decay tunnel. 
The beam dump at the end absorbs hadrons which did not decay.
Produced neutrinos with a peak energy of about 0.6 GeV are directed towards
a near detector placed at 280 m from the production target and 
the far detector, Super-Kamiokande, located 295 km away. 

At the first stage of the experiment, the T2K neutrino beamline was set up
to focus positively charged hadrons (the so-called ``positive'' focusing), 
to produce a $\nu_{\mu}$ beam. While charged pions generate most of 
the low energy neutrinos, charged kaons generate the high energy tail of 
the T2K beam, and contribute substantially to the intrinsic $\nu_{e}$
component in the T2K beam \cite{T2K_flux_paper}. 
An anti-neutrino beam can be produced by
reverting the current direction in the focusing elements of the beamline
in order to focus negatively charged particles (``negative'' focusing).

To calculate the neutrino flux {\sc FLUKA}\,\cite{FLUKA2008} 
and {\sc GEANT3} \cite{GEANT3} based simulation models are used.
The modeling of hadronic interactions is re-weighted using thin target hadron production
data, among which the main set have been provided by the NA61/SHINE experiment.
For the current analyzes of T2K,  uncertainties on the flux prediction at 
the $\nu_\mu$ peak energy are evaluated to be 12\% (see Fig.\,\ref{fig:Uncert}) and 2\%
for the absolute value of the flux  and 
for the far-to-near fluxes ratio, respectively. 

To reduce systematics related to hadron re-interactions inside the target
in future it is planned to use the T2K replica target results of NA61/SHINE
where hadron emission will be parametrized at the surface of the target.


\subsection{The NA61/SHINE experiment}

For the flux calculation  T2K  relies primarily on 
the hadron measurements performed by NA61/SHINE \cite{proposal_NA61} at CERN.
The NA61/SHINE apparatus \cite{NA61detector_paper} is a wide acceptance spectrometer 
at the CERN SPS. Most of detector components
were inherited from the NA49 experiment and are described in
\cite{NA49-NIM}. 
The NA61/SHINE spectrometer is built around five TPC detectors. 
Two of them are placed in the magnetic field produced by superconducting dipoles.
Particle identification is performed via measurements of energy losses in TPC
and time-of-flight measurements by a scintillator wall installed downstream of the spectrometer.

A 31 GeV/$c$ secondary hadron beam is produced from
400 GeV protons extracted from the SPS in slow extraction mode. 
%
%
For the cross section measurement NA61/SHINE used 
a thin graphite target (0.04 $\lambda_{I}$).
For the first physics analysis the pilot data collected in 2007 have been used.
The results on cross section of $\pi^\pm$ \cite{Abgrall:2011ae} 
and K$^+$ \cite{Abgrall:2011ts}
 have been integrated to the T2K beam simulation program so far.

\begin{figure*}[t!]
\centering
\includegraphics[width=0.44\textwidth]{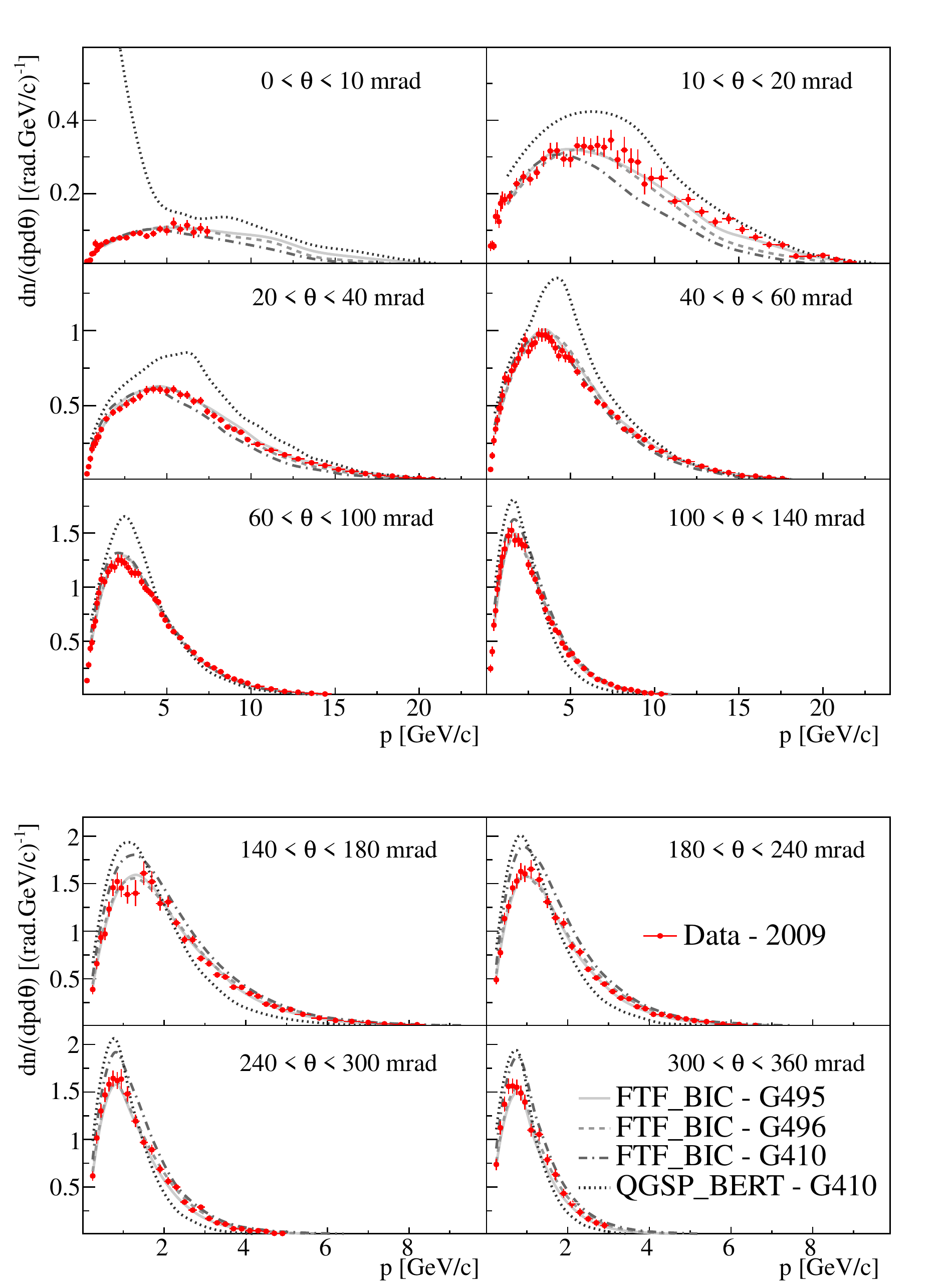}
\hspace{1cm}
\includegraphics[width=0.44\textwidth]{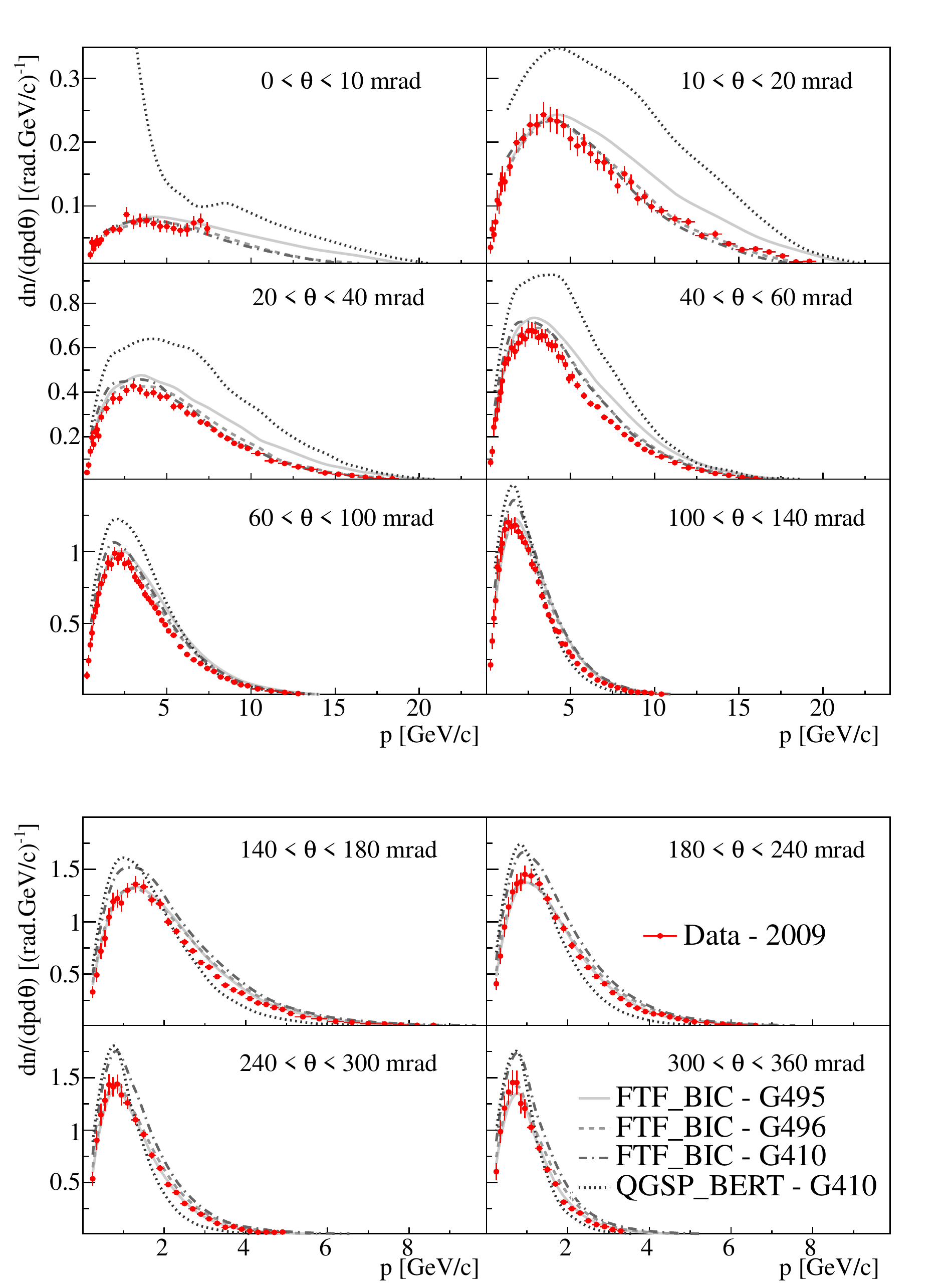}
\caption{Laboratory momentum distributions of the $\pi^+$ and $\pi^-$ multiplicities
produced in p-C interactions at 31 GeV/$c$ in different intervals of polar angle $\theta$
\cite{my_EPS13}.
Data points are overlapped by various GEANT4 model predictions \cite{GEANT4}.
}
\label{NA61_pions}
\end{figure*}

Although pilot data 2007 covered a significant part of the relevant hadron production 
phase space of T2K \cite{T2K_flux_paper} the statistical uncertainty was quite large.
In the year 2008 important changes have been introduced to the experimental 
setup of NA61/SHINE: new trigger logic, TPC read-out and DAQ upgrade,
additional sections of ToF wall, new beam-telescope detectors.
As a consequence of these upgrades the number of events recorded in 2009 and 2010
for about a same period of time have been increased by an order of magnitude
as compared to the 2007 run. This larger sample allows 
extraction 
of yields of $\pi^{\pm}$, K$^{\pm}$, K$^0_s$, $\Lambda$ and protons \cite{my_EPS13}
(for $\pi^{\pm}$ see Fig.\,\ref{NA61_pions}).
Furthermore, the phase space coverage 
of NA61/SHINE has been increased. 
Additional sections of ToF improve the acceptance 
at high $\theta$.
In the forward direction one profits from the use of the Gap TPC detector
which plays a key role in the analysis of forward produced particles.
The coverage of this kinematic domain is important for the muon monitor 
measurements of T2K \cite{T2K_NIM}. 
Statistics collected in 2007 and 2009 with the thin target is 
$0.7$ and $5.4$ millions of events, respectively. 

First physics results from the analysis of the full size T2K replica target (1.9 $\lambda_{I}$) data taken in 2007 have been published
\cite{NA61_LT}. 
A dedicated reconstruction method has been developed to provide results in a form that is of direct
interest for T2K. 
Yields of positively charged pions are reconstructed at the surface of the T2K replica
target in bins of the laboratory momentum and polar angle as a function of the longitudinal position
along the target. 
By parametrizing hadron yields on a surface of the target one constrains up to 90\% 
of the flux for both $\nu_\mu$ and $\nu_e$ components while only 60\% of neutrinos 
coming from decay of hadrons created in the primary interaction. 
Two methods (constraint of hadroproduction data at interaction vertices
and on a target surface) are consistent within their uncertainties achieved on statistics of 
the 2007 run.
The ultimate precision will come from the analysis of the T2K replica target 
\mbox{data} 2009 and 2010 \cite{Alexis}.

Following discussion initiated at NUFACT\,2011 a group from 7 US institutions expressed their 
interest in possibility of collecting data relevant for NuMI experiments 
(MINERvA, NOvA, MINOS+, MicroBooNE, ArgoNeut), 
and LBNE \cite{addendum_SPSC}. An important pilot run with a
proton beam of 120 GeV and a thin graphite target took place in July 2012. 
It resulted in 3.5 millions of recorded events.
An experience with these pilot data gives a basis to estimate an amount of efforts 
needed to fulfill the NuMI program in NA61/SHINE. 
A complete data taking is expected after the 2013/2014 shutdown of the CERN accelerator
complex. 
These measurements are in many senses analogous to the one 
presently performed by NA61/SHINE for T2K. 

\section{The NuMI beamline}

At NuMI beamline \cite{NuMI_beamline} protons of 120 GeV/$c$,
extracted from the main injector accelerator,
interact with a  90 cm long graphite target.
The latter is segmented longitudinally into 47 fins
soldered to water cooling line.
Particles which are produced in interactions are  focused 
by two magnetic horns. 
A significant fraction of mesons further decays in a 675 m long 
evacuated steel pipe and produces the neutrino beam. 
The pipe ends with a beam dump.
Ionization chamber upstream of  the absorber and muon station downstream 
are used to monitor parameters of the beam.
The energy spectrum of neutrinos could be adjusted by changing
the relative longitudinal position of the target with respect to the first horn:
the peak of the neutrino spectrum moves from 3 to 10 GeV 
by shifting the target from the nominal position inside the horn
by 2.5 m upstream.

At the earlier stage of the NuMI operation the neutrino beam was calculated
using {\sc FLUKA\,05} \cite{FLUKA2008} to describe interactions inside the target
and {\sc GEANT3} \cite{GEANT3} for a further particle propagation and decay.
A lack of hadron production data at NuMI energies 
makes the flux uncertainty
a dominant source of systematics in the neutrino flux prediction.
A significant disagreement between data and MC simulation
was observed in the analysis of
$\nu_\mu$ charged-current energy spectra measured by the Near Detector of MINOS 
\cite{MINOS_2008}.
Therefore these spectra were fitted to determine the pion and kaon yields 
off the NuMI target as a function of their transverse and longitudinal momenta.
%
To describe the hadron yields an analytic approximation was used. 
Obtained results were further applied to re-weight the original {\sc FLUKA\,05} distributions. 
For later analyzes of MINOS, in addition,
measurements by NA49 \cite{NA49_pC158} of the ratio of
$\pi^+$/$\pi^-$ yields were included as constraints in these fits \cite{MINOS_2011}.
This tuning procedure improves agreement between the
simulated ND energy spectrum and the data, however does not
significantly affect the predicted FD energy spectrum.

Nowadays NuMI experiments start using a {\sc GEANT4}\,\cite{GEANT4} based 
beamline simulation to predict the (anti)neutrino flux \cite{MINERvA_x}. 
Hadron production in the simulation was tuned to agree with the NA49 
measurements \cite{NA49_pC158}. 
{\sc FLUKA} 
is used to translate
NA49 measurements to proton energies between 12 and
120 GeV. Interactions not constrained by the
NA49 data are predicted using the FTFP hadron shower model.
In addition, ratios of pion and kaon cross sections released
by MIPP \cite{Lebedev} were included into analysis
which allowed to  reduce further the flux uncertainties. 
Values of fractional uncertainty calculated for the $\nu_\mu$ beam 
at MINERvA are shown in Fig.\,\ref{fig:Uncert}.

In addition to the traditional approach of re-weighting 
of hadron production yields at the interaction vertex,
recent results of MIPP on a direct measurements of particle production 
off a NuMI replica target \cite{MIPP} can be used in future.


\begin{figure*}[t!]
\centering
\includegraphics[width=0.397\textwidth]{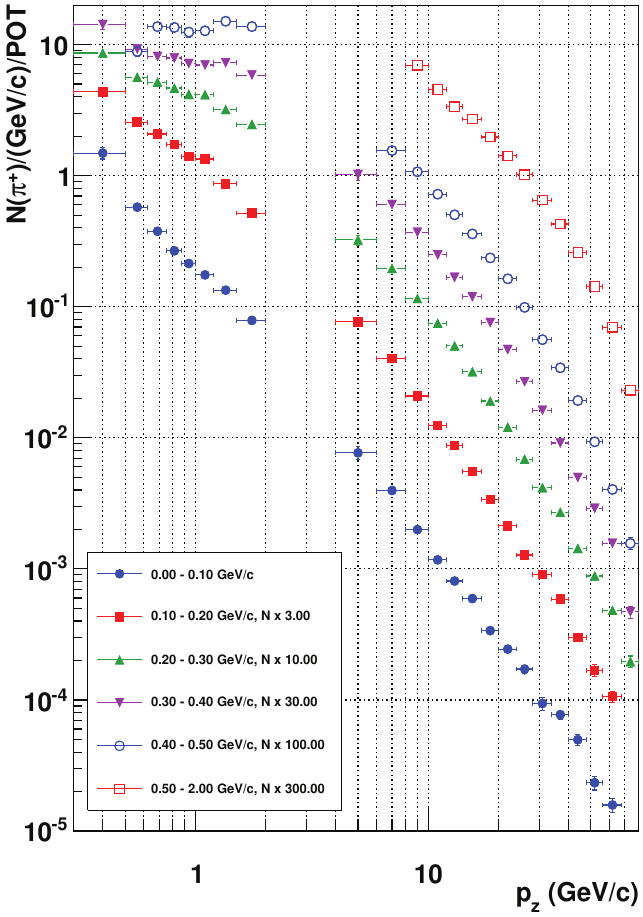}
\hspace*{1.5cm}
\includegraphics[width=0.392\textwidth]{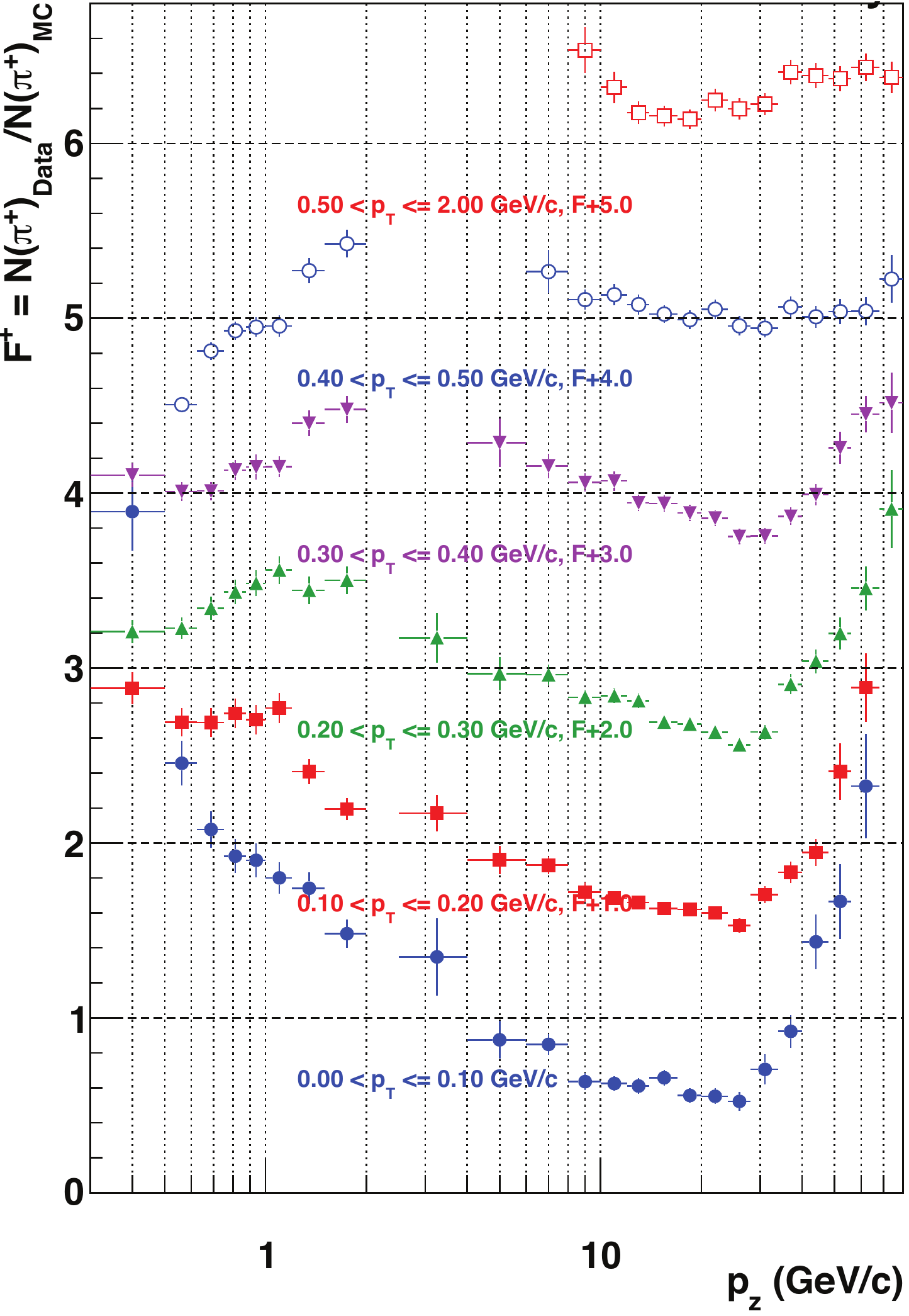}
\caption{Yields of $\pi^+$ \cite{MIPP} (left) and 
  ratios of data points to the GEANT4-based MC \cite{Paley_WC} (right) as a function 
  of $p_z$ in bins of $p_T$. 
  Different colors and markers represent bins of $p_T$, and the yields are scaled 
  and the ratios offset such that the points in different $p_T$ bins do not overlap. 
  Both statistical and systematic error bars are plotted.
}
\label{fig:MIPP_LT}
\end{figure*}

\subsection{The MIPP experiment}

The Main Injector Particle Production (MIPP) was a fixed target experiment 
\cite{MIPP} at Fermilab. 
The experiment was aimed to measure differential production yields of charged hadrons.
MIPP was taking data in 2004-2006 with a secondary proton beam of momenta 
from 5 to 85 GeV/$c$, or run with 120 GeV/$c$ proton beam from the Main Injector.
Various nuclear targets as well as a real full-size NuMI target have been used.
Track reconstruction for larger scattering angles is done by TPC 
which was placed inside the first dipole magnet.
Several stations of wire chambers are located along the beam axis of the spectrometer,
thus allow measurements of momentum up to 120 GeV/$c$. 
Particle identification is done via measurements of energy loss in TPC, 
time-of-flight and Cherenkov radiation.


First results of analysis of the MIPP data have been released in 2007 \cite{Lebedev}.
They were based on the analysis of 120 GeV/$c$ protons interacting with a thin carbon target.
Production ratios of $K^+$/$\pi^+$, $K^-$/$\pi^-$, $K^-$/$K^+$ and $\pi^-$/$\pi^+$ 
have been measured in 24 bins in longitudinal momentum from 20 to 90 GeV/$c$ and
transverse momentum up to 2 GeV/$c$. 
These ratios were compared to the ratios obtained from the fitted MINOS 
pion and kaon spectra on the NuMI target.
Reasonable agreement was found. 
In general, the precision of the MIPP measurement was limited by statistics
which that time was a limiting factor for understanding of the background.
Presently these results are used in the GEANT4 simulation of the NuMI beamline 
\cite{MINERvA_uncert_Harris}.


Recently new results on the inelastic cross section 
have been released \cite{MIPP_sig_prod}. 
They were obtained for 58 and 85 GeV/$c$ protons interacting with a LH$_2$ target, 
and 58 and 120 GeV/$c$ protons interacting with a carbon target.
Comparison of the cross section for the 58 GeV/$c$ beam with NA61/SHINE 
and other previous
measurements for the p-C interactions shows a reasonable agreement,
however overall uncertainty of the MIPP points is really large.


In addition to data collected with thin nuclear targets
MIPP also performed measurements of the hadron production yields 
off an actual NuMI target having 120 GeV/$c$ protons as a beam.
The goal was to verify the Monte Carlo calculation of 
hadron yields obtained with the traditional method of 
the cross section re-weighting at interaction vertices.


A measurement of $\pi^+$ and $\pi^-$ yields has been performed 
across 124 and 119 bins of longitudinal and transverse momentum,
respectively \cite{MIPP}.
Typical statistical and systematic uncertainties in most bins are between 5 and 10\%.
In contrast to the analysis of NA61/SHINE for T2K, 
the pion yields were integrated over bins along the target axis.
It still makes possible a direct comparison to the MC prediction
presently used by the NuMI experiments, 
however can introduce an additional uncertainty for 
interactions outside the target
if one decides to use this results for the re-weighting.
Comparison the MIPP spectra to GEANT4 based simulation 
is shown in Fig.\,\ref{fig:MIPP_LT}.
Data imply that MC tends to overestimate pion yields at higher momenta 
and underestimate at the focusing peak.
A similar tendency can be observed when data are compared 
to empirical parametrizations used in fits by MINOS \cite{Paley_WC}.
Understanding of these discrepancies should improve
our knowledge on hadron production and, in turn, 
decrease uncertainties of the neutrino flux 
in the NuMI beamline calculations.


\section{Conclusions}

The accelerator neutrino experiments that measure  interaction cross sections 
or perform   oscillation analyzes require a precise knowledge of 
the initial neutrino flux.
In the talk several approaches to constrain the neutrino flux have been presented. 
The application of these approaches for the T2K and NuMI beamlines has been discussed.
Importance of the hadron production data collected by NA61/SHINE and MIPP experiments 
have been emphasized.


{\small
{\setstretch{0}

}
}


\begin{thebibliography}{9}

\bibitem{review} S. E. Kopp, Phys.Rept. 439 (2007) 101
\bibitem{Bonesini}  M. Bonesini and A.Guglielmi, Phys. Rep. 433 (2006) 65

\bibitem{T2K_flux_paper} K. Abe {\em et al.}, Phys.Rev. D87 (2013) 012001

\bibitem{Boris} B. Popov, Nucl.Phys.Proc.Suppl. 235-236 (2013) 135

\bibitem{NA61_LT} N. Abgrall {\em et al.}, Nucl.Instrum.Meth. A701 (2013) 99

\bibitem{T2K_NIM} K. Abe {\em et al.}, Nucl.Instrum.Meth. A659 (2011) 106

\bibitem{NuMI_MuMon} D. Indurthy {\em et al.}, Conf.Proc. C0505161 (2005) 3892

\bibitem{Poster_0} Alysia Marino, proceedings of the Neutrino\,2014 conference
\bibitem{Poster_1} Jaewon Park, proceedings of the Neutrino\,2014  conference
\bibitem{Poster_2} Lu Ren, proceedings of the Neutrino\,2014  conference

\bibitem{Bodek} A. Bodek {\em et al.}, Eur.Phys.J. C72 (2012) 1973

\bibitem{NOMAD} V. Lyubushkin {\em et al.}, Eur.Phys.J. C63 (2009) 355

\bibitem{LBNE} C. Adams {\em et al.}, arXiv:1307.7335, April 2014

\bibitem{FLUKA2008} A. Fasso {\em et al.}, Report No. CERN-2005-10, 2005; \\
  G. Battistoni {\em et al.}, AIP Conf. Proc. 896, 31 (2007).

\bibitem{GEANT3} R. Brun {\em et al.}, CERN-W5013 (1994)

\bibitem{proposal_NA61} N. Antoniou {\it et al.}, CERN-SPSC-2006-034 

\bibitem{NA61detector_paper} N. Abgrall {\it et al.}, JINST 9 (2014) P06005
\bibitem{NA49-NIM} S. Afanasiev {\em et al.}, Nucl.Instrum.Meth. A430 (1999) 210

\bibitem{Abgrall:2011ae}  N. Abgrall {\it et al.}, Phys.Rev. C84 (2011) 034604
\bibitem{Abgrall:2011ts}  N. Abgrall {\it et al.}, Phys.Rev. C85 (2012) 035210




\bibitem{my_EPS13} A. Korzenev, talk at EPS-HEP2013, arXiv:1311.5719 

\bibitem{Alexis} A. Haesler, talk at 'Rencontres du Vietnam' in July 2014

\bibitem{addendum_SPSC} (US-NA61 collaboration) S. R. Johnson {\it et al.}, addendum to the NA61/SHINE proposal SPSC-P-330
	

\bibitem{GEANT4} S. Agostinelli {\em et al.}, Nucl.Instr.Meth. A506 (2003) 250;\\
J. Allison {\em et al.}, IEEE Transactions on Nuclear Science, 53(1), 270 (2006)

\bibitem{NuMI_beamline} S. E. Kopp, arXiv:0508001[physics.acc-ph] (2005)



\bibitem{MINERvA_uncert_Harris} D.Harris, presented at NuInt14 in May 2014





\bibitem{MINOS_2008} P. Adamson {\em et al.}, Phys.Rev. D77 (2008) 072002

\bibitem{NA49_pC158} C. Alt {\em et al.}, Eur. Phys. J. C49 (2007) 897

\bibitem{MINOS_2011} P. Adamson {\em et al.}, Phys.Rev.Lett. 107 (2011) 021801

\bibitem{MINERvA_x} L. Fields {\em et al.}, Phys.Rev.Lett. 111 (2013) 2, 022501

\bibitem{MIPP} J. Paley {\em et al.}, Phys.Rev. D90 (2014) 032001 

\bibitem{Lebedev}  A. Lebedev, PhD thesis, Harvard University, May 2007 

\bibitem{MIPP_sig_prod} S. Mahajan and R. Raja, arXiv:1311.2258 [hep-ex]

\bibitem{Paley_WC} J. Paley, FNAL JETP seminar, Apr 8, 2014


\end{thebibliography}
\end{document}